\documentclass[
reprint,
superscriptaddress,
showpacs,
 amsmath,amssymb,
 aps,
prl,
longbibliography,
]{revtex4-1}

\usepackage{graphicx}
\usepackage{amsmath}
\usepackage{mathrsfs}
\usepackage{multirow}

\begin{document}
\title{Strong Modification of Radiative Transition Rates\\
due to a Breit-Interaction-Induced Avoided Crossing}


\author{Zhimin Hu}
\affiliation{Laser Fusion Research Center, China Academy of Engineering Physics, Mianyang 621900, China}

\author{Gang Xiong}
\affiliation{Laser Fusion Research Center, China Academy of Engineering Physics, Mianyang 621900, China}

\author{Xiang Gao}
\affiliation{Beijing Computational Science Research Center, Beijing 100084, China}
\affiliation{Institute for Theoretical Physics, Vienna University of Technology, A-1040 Vienna, Austria, EU}

\author{Nobuyuki Nakamura}
\affiliation{Institute for Laser Science, The University of Electro-Communications, Tokyo 182-8585, Japan}

\author{Ke Yao}
\affiliation{Shanghai-EBIT Laboratory, Key Laboratory of Nuclear Physics and Ion-beam Application (MOE), Fudan University, Shanghai 200433, China}

\author{Chengsheng Wu}
\affiliation{Beijing Computational Science Research Center, Beijing 100084, China}
\affiliation{Department of Engineering Physics, Tsinghua University, Beijing 100084, China}

\author{Naoki Numadate}
\affiliation{Institute for Laser Science, The University of Electro-Communications, Tokyo 182-8585, Japan}

\author{Chengwu Huang}
\affiliation{Laser Fusion Research Center, China Academy of Engineering Physics, Mianyang 621900, China}

\author{Yulong Ma}
\affiliation{Institute for Applied Physics and Computational Mathematics, Beijing 100088, China}

\author{Yong Wu}
\affiliation{Institute for Applied Physics and Computational Mathematics, Beijing 100088, China}

\author{Yueming Li}
\affiliation{Institute for Applied Physics and Computational Mathematics, Beijing 100088, China}

\author{Yaming Zou}
\affiliation{Shanghai-EBIT Laboratory, Key Laboratory of Nuclear Physics and Ion-beam Application (MOE), Fudan University, Shanghai 200433, China}

\author{Baohan Zhang}
\affiliation{Laser Fusion Research Center, China Academy of Engineering Physics, Mianyang 621900, China}

\author{Jiamin Yang}
\affiliation{Laser Fusion Research Center, China Academy of Engineering Physics, Mianyang 621900, China}

\date{\today}

\begin{abstract}
We present the observations of x-rays emitted from the $1s2s^{2}2p_{1/2}2p_{3/2}$ inner shell excited state of B-like W and Bi ions.
The relative transition rates are obtained for two dominant radiative transitions to $1s^{2}2s^{2}2p_{1/2}$ and $1s^{2}2s^{2}2p_{3/2}$.
The experimental results and the comparison with rigorous relativistic calculations show that the rates of the strong electric dipole allowed $1s^22s^22p$ -- $1s2s^22p^2$ transitions are strongly modified due to a drastic change in the wavefunction caused by the Breit interaction.
\end{abstract}
\maketitle

The relativistic and quantum electrodynamics (QED) effects in electron--electron interaction can be described by the Breit interaction, which was first introduced to precisely describe the energy levels of neutral He by G.~Breit around 1930~\cite{PhysRev.36.383}.
It includes magnetic interaction consisting of spin--spin, orbit--orbit, and spin--other-orbit interactions, and retardation in the exchange of a virtual photon between the interacting electrons.
Recently, it has been found that the Breit interaction can have a relatively large or even dominant contribution to collision processes of highly charged heavy ions with electrons, such as electron-impact ionization~\cite{ISI:A1994NU48700007}, electron-impact excitation~\cite{ISI:000319277300011,PhysRevA.99.032706}, dielectronic recombination (DR)~\cite{ISI:000253336900025,PhysRevLett.108.073002,ISI:000362343200003,ISI:000353030800003,Tong2015,Nakamura2016,PhysRevA.83.020701}, and resonant transfer and excitation~\cite{PhysRevA.68.042712}.

On the other hand, the Breit interaction has a relatively small contribution to the atomic structure and radiative processes.
For the radiative transition from the level $\left| a \right>$ to $\left| b \right>$, the transition rate is determined by the transition frequency $\omega$ and the transition matrix $\left< b \right| \hat{o} \left| a \right>$, where $\hat{o}$ represents the operator for the radiative transition.
The Breit interaction has no contribution to the operator $\hat{o}$ and only a small contribution to $\omega$ and the wavefunctions of $\left| a \right>$ and $\left| b \right>$.
Consequently, transition rates are less affected by the Breit interaction. 
Exceptions can only be found in very limited cases 
where the transition rate is very small (usually lower than $10^3$ s$^{-1}$) and hence is very sensitive to the subtle change of the wavefunction~\cite{PhysRevLett.78.4355,PhysRevA.64.042507,PhysRevLett.95.183001,Han2012}.
For example, for the weak spin-forbidden inter-combination transition $2s^2\:^1\!S_0$ -- $2s2p\:^3\!P_1$ of C$^{2+}$ with a rate of about $10^2$ s$^{-1}$, a relatively large modification of the transition rate was previously reported due to the enhancement of the mixing between the metastable $2s2p\:^3\!P_1$ state and the $2s2p\:^1\!P_1$ excited state induced by the Breit interaction~\cite{PhysRevLett.78.4355,PhysRevA.64.042507}.
A similar degree of modification due to the Breit interaction was also found in higher multipole transitions, such as the magnetic dipole or electric quadrupole transitions among the electronic ground configurations in Ar$^{13+}$~\cite{PhysRevLett.95.183001} and O$^+$~\cite{Han2012} ions. 

In this Letter, we report experimental observations of the $1s^22s^22p$--$1s2s^22p^2$ transitions in B-like heavy ions.
In contrast to the previous studies described above~\cite{PhysRevLett.78.4355,PhysRevA.64.042507,PhysRevLett.95.183001,Han2012}, the rates of the strong electric dipole ($E1$) allowed transitions ($\sim10^{16}$ s$^{-1}$) are shown to be substantially modified by the Breit interaction.
It is due to the drastic change in the wavefunctions of the $1s2s^22p^2$ state resulting from the relativistic electron correlation effect appeared as a Breit-interaction-induced avoided crossing.

In order to produce and study inner shell excited states of highly charged heavy ions, an electron beam ion trap (EBIT)~\cite{Marrs1,Knapp2} is an effective device.
An EBIT consists of a Penning-like ion trap and a high-energy, high-density electron beam traveling through the trap.
Highly charged ions (HCIs) are produced by successive electron impact ionization of the trapped ions.
Inner shell excited states of B-like ions can be produced by electron-impact inner shell excitation of B-like ions or inner shell ionization of C-like ions.
However, resonant dielectronic capture (DC) is the most efficient method to state-selectively produce inner shell excited states.
In the DC process of Be-like ions, a free electron is captured by the Be-like ion while exciting an inner shell electron so that an inner shell excited B-like ion is produced.
Owing to the quasi-monoenergetic nature of the electron beam in an EBIT (the electron energy width is typically about 50~eV), an inner shell excited state can be produced state-selectively by tuning the electron energy at the resonance energy.
The inner shell excited state can be stabilized either by radiative decay or by autoionization, but the branching ratio of the radiative decay is almost unity for highly charged heavy ions.
The two-step process consisting of DC and radiative decay is referred to as DR.
Thus, in the present study, the radiative transitions of inner shell excited B-like ions were studied by observing x-rays arising from DR into Be-like ions.

\begin{figure}
 \includegraphics[width=0.5\textwidth]{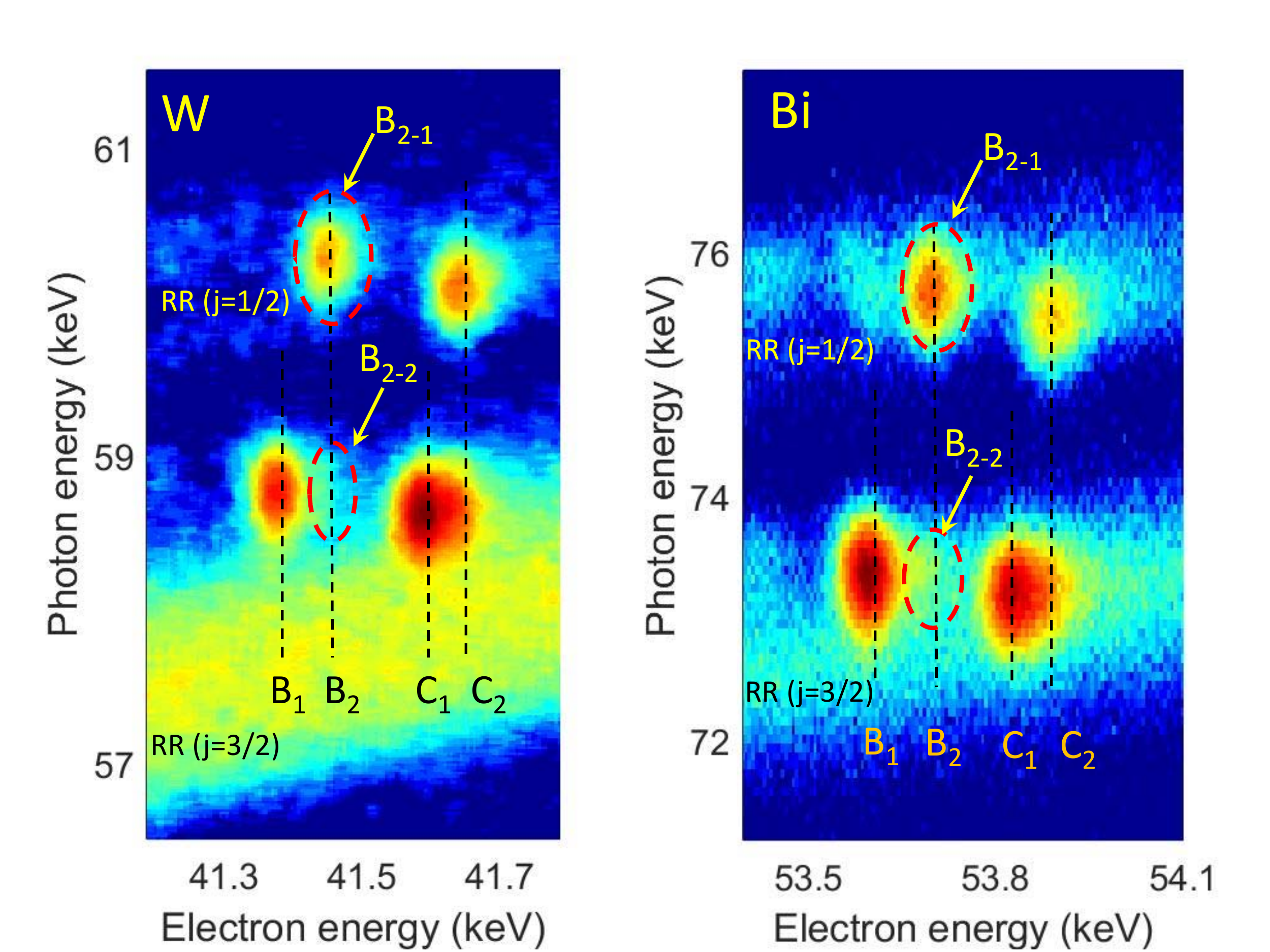}%
 \caption{\label{fig1}(color online). Two dimensional plots of the x-ray spectra obtained as a function of electron energy at the $KL_{12}L_{3}$ DR region (left for W and right for Bi).
 X-ray counts are plotted as a brightness of color.
 Bright spots correspond to x-ray enhancement due to DR.
 B$_1$ and B$_2$ are DR forming B-like ions via $[1s2s^22p_{1/2}2p_{3/2}]_{5/2}$ and $[1s2s^22p_{1/2}2p_{3/2}]_{3/2}$, respectively.
 C$_1$ and C$_2$ are DR forming C-like ions via $[1s2s^22p_{1/2}^22p_{3/2}]_J$ with $J$=1 and 2, respectively.
 }
 \end{figure}

In the present experiments, the upgraded Shanghai-EBIT~\cite{ISI:000342910500014} and the Tokyo-EBIT~\cite{Nakamura1997} were used for W and Bi, respectively.
The volatile organic compound $\rm W(CO)_{6}$ was used for W injection into the trap region of the Shanghai-EBIT~\cite{PhysRevA.91.060502}, and an effusion cell~\cite{YamadaRSI} was used for Bi injection into the Tokyo-EBIT.
The operation parameters, such as beam current and trapping potential, were optimized to obtain the maximum fraction of the Be-like ions in the trap region.
To produce inner shell excited states of B-like ions, the electron beam energy was swept over the resonance energy region for $KL_{12}L_{3}$ DR into Be-like ions.
$KLL$ denotes the process where the incident electron is captured into the $L$ shell while the $K$ shell bound electron is excited to the $L$ shell.
$L_{12}$ and $L_3$ denote the orbitals with $j=1/2$ ($2s_{1/2}$ or $2p_{1/2}$) and that with $j=3/2$ ($2p_{3/2}$).
X-ray photons were detected by an ORTEC high-purity germanium detector placed in the direction perpendicular to the electron beam.
The x-ray energy and the electron beam energy were simultaneously registered by a multi-channel event-mode data acquisition system which was triggered by each x-ray photon.

The recorded two-dimensional (2D) x-ray spectra are shown in Fig.~\ref{fig1}.
It took approximately 230 hours for W and 60 hours for Bi to achieve enough statistics.
On the diagonal lines arising from radiative recombination (RR), x-ray enhancement due to DR was observed at mainly four resonance energies for both W and Bi.
Two of them (denoted as B$_1$ and B$_2$) at the lower energy side correspond to DR forming B-like from Be-like ions, and another two (C$_1$ and C$_2$) at the higher energy side correspond to that forming C-like from B-like ions as indicated in the figure.
The intermediate inner shell excited state for B$_1$ has a total angular momentum of $J=5/2$, and can be well described by a single $jj$-coupled configuration state  $[(1s2s^22p_{1/2})_12p_{3/2}]_{5/2}$.
On the other hand, the intermediate state for B$_2$ has $J=3/2$, and it can be predominantly represented as a linear combination of the two $jj$-coupled configurations with $J=3/2$, which are $[(1s2s^22p_{1/2})_02p_{3/2}]_{3/2}$ and $[(1s2s^22p_{1/2})_12p_{3/2}]_{3/2}$.
There are thus two states arising from the linear combination as follows.
\begin{eqnarray}
\left| + \right> = c_1 \left| (1s2p_{1/2})_0 2p_{3/2} \right> + c_2 \left| (1s2p_{1/2})_1 2p_{3/2} \right> , \label{eq:p}\\
\left| - \right> = c_2 \left| (1s2p_{1/2})_0 2p_{3/2} \right> - c_1 \left| (1s2p_{1/2})_1 2p_{3/2} \right> , \label{eq:n}
\end{eqnarray}
where $c_1$ and $c_2$ are the mixing coefficients that satisfy $c_1^2+c_2^2=1$.
It is noted that $2s^2$ is omitted from the wavefunction notations used in the above equations. 
In the atomic number ($Z$) region of the present interest, these two $jj$-coupled configurations strongly interact.
In this strong correlated regime, it is very interesting to note that the total spin of the $\left| + \right>$ state is mainly 1/2, whereas that of the $\left| - \right>$ state is mainly 3/2~\cite{Ref19}.
The energy separation between these two levels is within the present experimental electron beam resolution.
Nevertheless, according to the approximate spin conservation in the electron collision with the ground $1s^22s^2$ closed shell Be-like ions, the $\left| + \right>$ state with a total spin of 1/2 is nearly exclusively populated.
Therefore, the intermediate state of the B$_2$ resonance corresponds to $\left| + \right>$~\cite{Ref19}.

\begin{figure}
 \centering
 \includegraphics[width=0.5\textwidth]{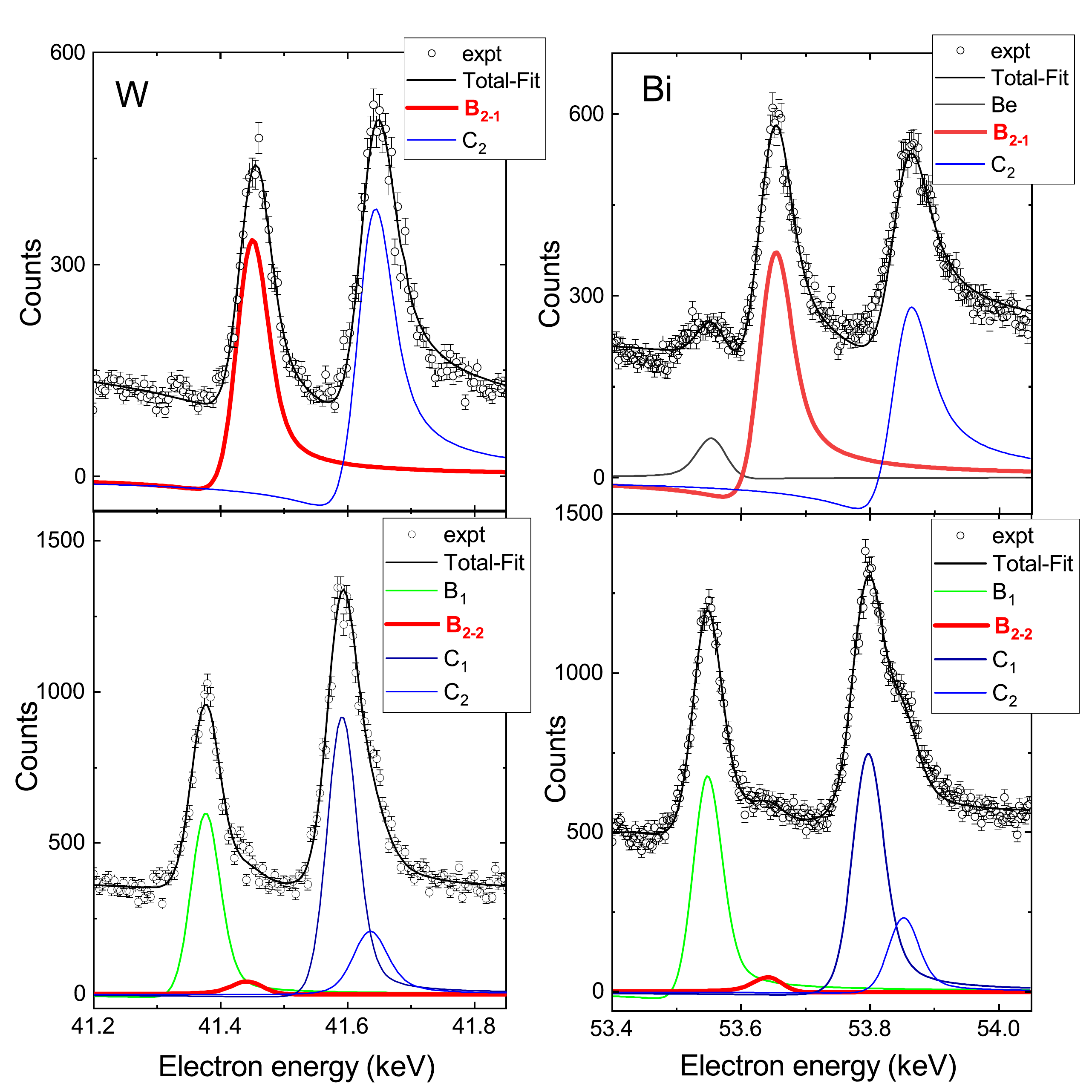}%
\caption{\label{fig2}(color online). 
Black circles represent the observed X-ray counts plotted as a function of electron energy for the regions of RR into the $2l_{1/2}$ (upper) and $2l_{3/2}$ (lower) orbitals (left for W and right for Bi).
Solid lines represent the DR resonance profiles fitted to the observed x-ray counts.
B$_{2-1}$ and B$_{2-2}$ are the transitions of the present interest, corresponding to the radiative decay from $[1s2s^22p_{1/2}2p_{3/2}]_{3/2}$ to $1s^{2}2s^{2}2p_{1/2}$ and $1s^{2}2s^{2}2p_{3/2}$, respectively.
}
\end{figure}

\begin{table*} 
 \caption{\label{tab1}
 Rates $A$ for the radiative transitions from the $1s2s^22p_{1/2}2p_{3/2}$ $\left| + \right>$ level defined in Eq.~(\ref{eq:p}) to $1s^22s^22p_{1/2}$ ($A_1$) and to $1s^22s^22p_{3/2}$ ($A_2$) and their ratio $R=A_1/A_2$.
 The superscript ``th" denotes the theoretical values whereas ``exp" the experimental ones.
 Two theoretical values are given: one (C) obtained with the Coulomb interaction only and another (C+B) obtained with including the Breit interaction.
  }
 \begin{ruledtabular}
 \begin{tabular}{cccccccc}
 & Transition & Label & \multicolumn{2}{c}{$A^{\rm th}$ ($10^{16}$ s$^{-1}$)} & \multicolumn{2}{c}{$R^{\rm th}$} & $R^{\rm exp}$\\ 
\cline{4-5}\cline{6-7}
& & & C &  C+B & C  &C+B  &  \\
\hline
W ($Z=74$)	&transition 1 ($\to1s^22s^22p_{1/2}$)& $A_1$ & 2.0 & 1.8 & \multirow{2}{*}{2.5} & \multirow{2}{*}{4.7} & \multirow{2}{*}{5.1$\pm1.3$}\\
& transition 2 ($\to1s^22s^22p_{3/2}$) & $A_2$ & 0.81 & 0.38 & & & \\

Bi ($Z=83$)	& transition 1 ($\to1s^22s^22p_{1/2}$) & $A_1$ & 3.0 & 2.5 & \multirow{2}{*}{2.3} & \multirow{2}{*}{5.7} & \multirow{2}{*}{5.8$\pm1.2$} \\
& transition 2 ($\to1s^22s^22p_{3/2}$) & $A_2$ & 1.3 & 0.43 & & & \\
 \end{tabular}
 \end{ruledtabular}
 \end{table*}

\begin{figure}
 \includegraphics[width=0.5\textwidth]{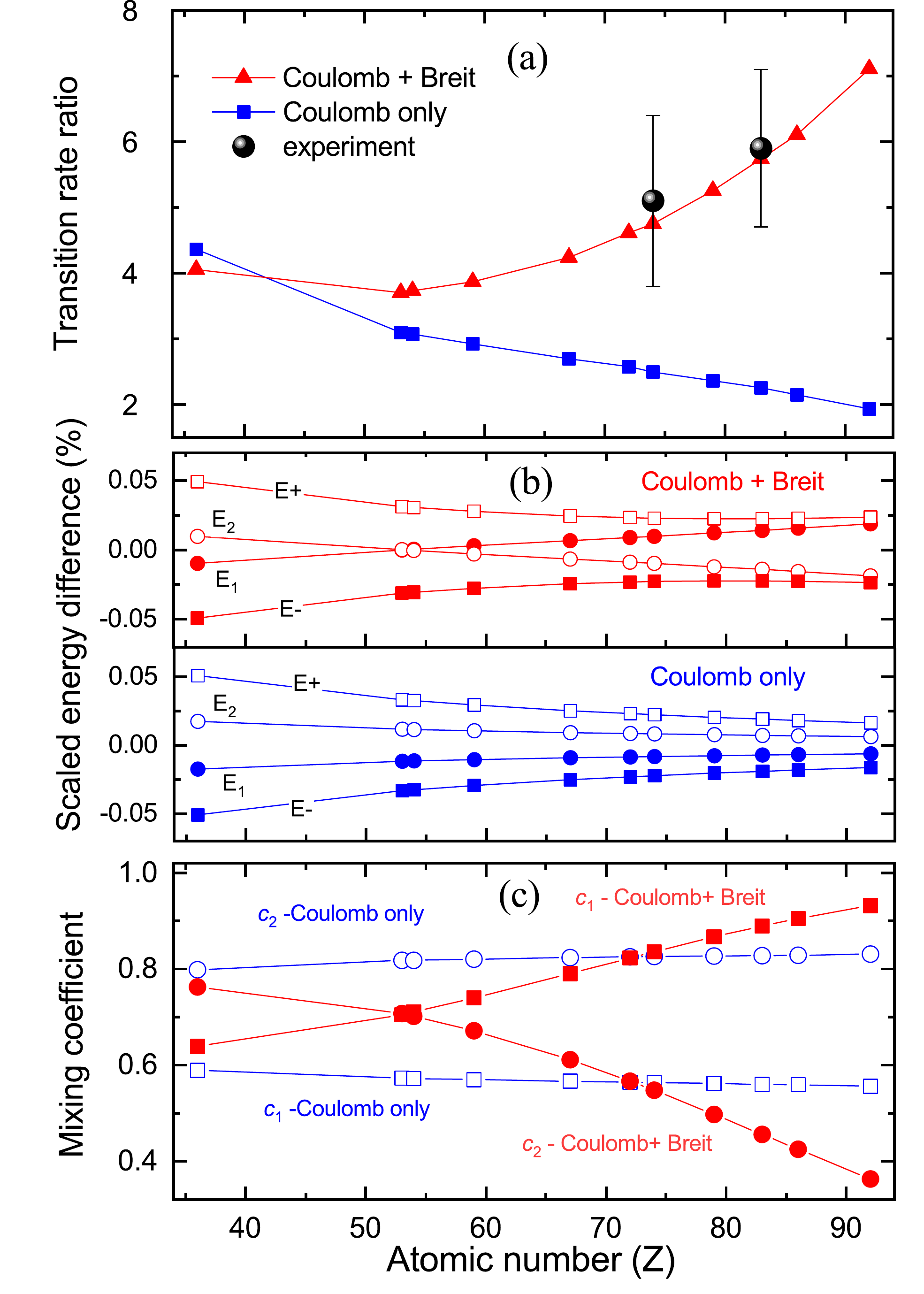}%
 \caption{\label{fig3}(color online).
 (a) Transition rate ratio between two radiative transition paths of the $\left| + \right>$ level.
 The solid lines with triangles and squares represent the theoretical values obtained with and without the Breit interaction, respectively.
 The black circles represent the experimental results.
 (b) Scaled energy difference defined by $(E-E_{\rm av})/E_{\rm av}$, where $E_{\rm av}$ is the averaged energy.
 $E_+$ and $E_-$ represent the energy of the $\left| + \right>$ and $\left| - \right>$ levels, respectively.
 $E_1$ and $E_2$ represent the energy of the $jj$-coupled configurations $\left| (1s2p_{1/2})_0 2p_{3/2} \right>$ and $\left| (1s2p_{1/2})_1 2p_{3/2} \right>$ levels, respectively (see Eqs.~(\ref{eq:p}) and (\ref{eq:n})).
 (c) Mixing coefficients defined in Eqs.~(\ref{eq:p}) and (\ref{eq:n}).
}
 \end{figure}

Here we investigate the radiative decay process of the level $\left| + \right>$, which has two $E1$ allowed decay paths to $1s^{2}2s^{2}2p_{1/2}$ (transition 1) and $1s^{2}2s^{2}2p_{3/2}$ (transition 2), which correspond to B$_{\rm 2\mathchar`-1}$ and B$_{\rm 2\mathchar`-2}$ in Fig.~\ref{fig1}, respectively.
As confirmed in Fig.~\ref{fig1}, the x-ray intensity of transition 1 (B$_{\rm 2\mathchar`-1}$) is much stronger than that of transition 2 (B$_{\rm 2\mathchar`-2}$).
In order to obtain experimental x-ray intensity, DR resonance profiles were fitted to the electron energy dependence of the x-ray counts obtained by projecting the 2D plot in Fig.~\ref{fig1} onto the electron beam energy axis as shown in Fig.~\ref{fig2}.
The neighboring DR forming Be-like and C-like ions were also included in the fitting procedure.
Fano profile functions were used for the fitting as strong quantum interference between DR and RR is known to exist for these resonances~\cite{PhysRevLett.94.203201,ISI:000372398300007}.
Since the experimental x-ray intensity $I$ is determined by $n\cdot A\cdot W(90^\circ)$ ($n$ is the population of the upper level, $A$ the transition rate, and $W(\theta)$ the angular distribution factor), the radiative transition rate ratio $R=A_1/A_2$ between transitions 1 and 2, which have the identical upper level, can be obtained by
\begin{equation}\label{eq1}
R=\frac{I_{\rm B_{\rm 2\mathchar`-1}} \cdot W_{\rm B_{\rm 2\mathchar`-2}}(90^\circ)}{I_{\rm B_{\rm 2\mathchar`-2}} \cdot W_{\rm B_{\rm 2\mathchar`-1}}(90^\circ)}.
\end{equation}
The ratio determined from the experimental intensity are listed in Table~\ref{tab1} and plotted in Fig.~\ref{fig3}(a).
Angular distribution factors calculated by the relativistic configuration interaction approach~\cite{ISI:000378427900088} were used to obtain the ratio.

For comparison, theoretical calculations of radiative transition rates were done by using relativistic wavefunctions obtained in the multi-configuration Dirac-Fock approach (details will be given elsewhere~\cite{Ref19}) with the Dirac-Coulomb-Breit Hamiltonian, which can be written as
\begin{equation}\label{eq3}
  H_{\rm DCB}=\sum^{N}_{i}\left[c\alpha_i\dot p_i+(\beta_i-1)c^2 - V(r_i) \right]+ \sum^{N}_{i<j} V^{\rm ee}_{ij},
\end{equation}
where $c$ is the speed of light, $\alpha$ and $\beta$ (4$\times$4) Dirac matrices, $V(r_i)$ the potential of nuclear charge, and $V^{\rm ee}$ the electron-electron interaction potential.
Theoretically obtained transition rates and ratios are also listed in Table~\ref{tab1} and plotted in Fig.~\ref{fig3}.
We made calculations with and without including the Breit interaction in $V^{\rm ee}$.
As shown in Fig.~\ref{fig3}(a), the transition rate ratio monotonously decreases as $Z$ increases when only the Coulomb interaction is considered in $V^{\rm ee}$.
On the other hand, the ratio obtained with including the Breit interaction shows a completely different behavior, which reproduces the experimental results quite well.
This is due to the strong Breit interaction effect especially on the rate of transition 2, which is decreased by a factor of two for W and three for Bi, compared with the rates obtained without the Breit interaction as shown in Table~\ref{tab1}.
In general, the Breit interaction is considered to have little effect on the wavefunctions and radiative transition rates because the electron--electron interaction is regarded as a minor effect compared with the dominant central Coulomb field in highly charged heavy ions.
However, this strong Coulomb interaction can squeeze the intra-shell orbitals close to each other.
It is thus reasonable to expect that the states composed of those orbitals may interact strongly while electron correlations play very important roles in those states.

In order to consider the underlying mechanism that realizes this strong modification on the radiative transition rates due to the Breit interaction, the energy of the $\left| +\right>$ and $\left| -\right>$ levels and the mixing coefficients $c_1$ and $c_2$ are plotted in Fig.~\ref{fig3}(b) and (c), respectively.
As understood from the comparison between the upper and the lower panel of Fig.~\ref{fig3}(b), the Breit interaction has little effect on the energy of the $\left| +\right>$ and the $\left| -\right>$ levels.
The effect is as small as in the order of 0.01\% or less.
This is consistent with the general understanding that the energy levels are little affected by the Breit interaction.
On the other hand, however, the mixing coefficients are greatly modified by the Breit interaction as shown in Fig.~\ref{fig3}(c).
The coefficient $c_2$ is always larger than $c_1$, i.e. the main $jj$-coupled component of the $\left| +\right>$ level is $\left| (1s2p_{1/2})_1 2p_{3/2} \right>$ over a whole range of $Z$ shown in Fig.~\ref{fig3} when the Breit interaction is not considered.
However, when the Breit interaction is considered, the magnitude between the coefficients $c_1$ and $c_2$ is inverted at $Z=54$ and $c_1$ dominates over $c_2$ at higher $Z$.
Thus the nature of the wavefunction of $\left| +\right>$ drastically changes from $\left| (1s2p_{1/2})_1 2p_{3/2} \right>$-like to $\left| (1s2p_{1/2})_0 2p_{3/2} \right>$-like wavefunction as $Z$ increases due to the Breit interaction effect.
In the consideration of the $jj$-coupled scheme, $[(1s2s^22p_{1/2})_1 2p_{3/2}]_{J=3/2}$ can decay both to $1s^22s^22p_{1/2}$ and to $1s^22s^22p_{3/2}$ whereas $[(1s2s^22p_{1/2})_0 2p_{3/2}]_{J=3/2}$ can decay only to $1s^22s^22p_{1/2}$.
Consequently, the ratio of transition 1 to transition 2 increases with $Z$ in the decay of the $\left| +\right>$ level.

As shown in the bottom panel of Fig.~\ref{fig3}(b), the single configuration energies of $\left| (1s2p_{1/2})_0 2p_{3/2} \right>$ and $\left| (1s2p_{1/2})_1 2p_{3/2} \right>$ ($E_1$ and $E_2$) are close to each other but never cross when the Breit interaction is not considered.
Thus the $\left| +\right>$ and $\left| -\right>$ levels are avoiding through the strong configuration mixing but the nature of the wavefunction is not changed over the $Z$ range shown in the figure as confirmed from the almost flat dependence of the mixing coefficients on $Z$ (open symbols in Fig.~\ref{fig3}(c)).
On the other hand, the Breit interaction induces the crossing between the single configuration energies as confirmed in the upper panel of Fig.~\ref{fig3}(b), which results in the strongly coupled avoided crossing between the $\left| +\right>$ and $\left| -\right>$ levels and the drastic change in the wavefunction before and after the crossing.
This Breit interaction induced avoided crossing is responsible for the significant modification on the radiative transition rates observed in the present experiments.

Avoided crossings appeared in the $Z$ dependence of the energy levels were also studied in the previous studies~\cite{Nakamura2000,Beiersdorfer2016}.
They studied the crossing realized by the change in the energy level ordering due to the coupling scheme change from $LS$ to $jj$.
Thus the level crossing mechanism is essentially different from that in the present study.
In addition, it is worth emphasizing that the most distinguished difference is in the role of the electron--electron interaction.
In the previously studied system~\cite{Nakamura2000,Beiersdorfer2016}, the electron correlation effect can be viewed as a perturbation, and the most prominent effects appear just in the ions around the level crossings.
On the other hand, the presently studied ions belong to a strongly correlated system, where the effects last over a whole range of $Z$ shown in Fig.~\ref{fig3}.
As the intra-shell electron correlations should be important in many HCIs, the present study should have great implications in HCI-related applications.

In summary, the electric dipole allowed radiative decays from the strongly correlated $1s2s^{2}2p^{2}$ inner shell excited state of B-like heavy ions have been experimentally studied with the electron beam ion traps in Shanghai and Tokyo.
By comparing the experimental data with rigorous calculations, the strong Breit interaction effect on the radiative transition rates has been shown for the first time.
The present study opens a new aspect of understanding the transition dynamics of few-electron excited ions, in particular, for the atomic states in strong Coulomb fields of highly charged heavy ions~\cite{RevModPhys.90.045005}.
This work is also beneficial for the x-ray plasma spectroscopy.
Inner shell excited states due to dielectronic capture into highly charged ions play an important role in the formation of dielectronic satellite x-ray lines, which are intensively used to infer the temperature and density of hot astrophysical and fusion device plasmas~\cite{Porquet2010,PhysRevLett.71.1007,ISI:A1995QD12400136,PhysRevLett.42.1606}. 
Radiative transitions of inner shell excited states should thus be understood in detail for accurate plasma diagnostics.

\begin{acknowledgments}
Z. Hu and X. Gao contributed equally to this work.
We would like to acknowledge Prof. Tu-Nan Chang at University of Southern California for helpful discussions.
This work was supported by the CAEP foundation (Grant Nos. CX2019022 and YZJJLX2017010) and the National Nature Science Foundation of China (No.11675158).
X. Gao thanks the support by the National Natural Science Foundation of China (Grant Nos.11774023 and U1930402), the National Key Research and Development Program of China (Grant No.2016YFA0302104), the National High-Tech ICF Committee in China.
K. Yao thanks the support by National Natural Science Foundation of China (Grant No.11874008) and the National Key Research and Development Program of China under Grant No. 2017YFA0402300.
\end{acknowledgments}

\bibliography{reference}

\end{document}